\documentstyle[aps,twocolumn,epsfig]{revtex}
\begin{document}
\newcommand{\ve}[1]{\mbox{\boldmath $#1$}}
\twocolumn[\hsize\textwidth\columnwidth\hsize
\csname@twocolumnfalse%
\endcsname

\draft

\title {The Yrast Spectra of Weakly Interacting
Bose-Einstein Condensates}
\author{B. Mottelson}
\address{Nordita, Blegdamsvej 17, DK-2100 Copenhagen {\O},\\ Denmark}
\date{\today}
\maketitle
 
\begin{abstract}

The low energy quantal spectrum is considered as a function of the total angular momentum 
for a system of weakly interacting bosonic atoms held together by an external
isotropic harmonic potential. It is found that besides the usual condensation
into the lowest state of the oscillator, the system exhibits two additional
kinds of condensate and associated thermodynamic phase transitions. These new
phenomena are derived from the degrees of freedom of ``partition space" which
describes the multitude of different ways in which the angular momentum can be
distributed among the atoms while remaining all the time in the lowest state
of the oscillator.

\end{abstract}
\pacs{PACS numbers: 03.75.Fi, 05.30.Jp, 67.40.Db}
\vskip2pc] 
 
The properties of the non-zero angular momentum states of Bose-condensates
of atoms in external traps have been addressed in several recent 
publications \cite{stringari}. Special interest attaches to the lowest energy 
quantum states with a
given angular momentum. Borrowing from nuclear physics terminology we shall
refer to this part of the quantum spectrum as the yrast region [2].
Theoretical work on the yrast region for Bose condensates has, with the
exception of \cite{wilkin}, been based on the Gross-Pitaevski mean field
approximation \cite{rokhar}. In the present Letter the yrast spectra are
described in terms of the elementary modes that carry the angular momentum. It
is found that this approach reveals structures in the yrast spectra that were
not recognized in previous work.   The picture of the yrast region described in
this note  draws heavily on ideas developed in collaboration with A. Bohr
\cite{bm} in  connection 
with the analysis of the stability of persistent currents in superconductors.
 
The model considered, which has previously been discussed by Wilkin et 
al.\cite{wilkin}, involves $N$ spin zero bosons moving in an
isotropic harmonic confining potential and a contact interaction with a
strength that is assumed to be weak, in the sense:
\begin{equation}
v N \ll \hbar \omega_0,
\end{equation}
where $v$ is the expectation value of the interaction between two particles
both in the ground state of the oscillator, and $\hbar \omega_o$ is the
quantum energy of the confining potential.  Although the so-far published 
experiments have mainly studied systems that do
not satisfy the condition (1), systems that do satisfy this condition are
accessible with current experimental techniques and would be especially
interesting since this implies that the coherence length in the atomic cloud
becomes larger than the size of the system, bringing into sharpest focus the
mesoscopic nature of these systems.   The eigenstates of a single
particle moving in an harmonic oscillator can be labeled by the quantum
numbers $(n l m)$ where $n$ is the radial quantum number, $l$ is the angular
momentum, and $m$ is the component of $l$ on the axis of quantization. The
energy of such a state is $\epsilon = (2n + l + 3/2) \hbar \omega_0$, and for
the yrast states we need only consider occupancy of states with $n = 0$ $m = l
\geq 0.$ (Without loss of generality we assume that the total angular momentum
is in a state of maximum alignment $L = M$). Thus the relevant single particle
states involve only a single non-negative quantum number, $m$.\\ \indent In
the yrast state with total angular momentum $\hbar L$, the motion in the
oscillator will contribute an excitation energy $L \hbar \omega_0$ with
respect to the ground state. However, this configuration is degenerate in  the
absence of the interactions, since the oscillator energy is the same whether
the $L$ quanta are distributed among the particles with one quantum to each of
$L$ different particles, or with 2 quanta on one particle and one quantum on 
each of  $L-2$ other particles etc. Since the multiplicity $p(L)$ of the
partitions of the integer $L$ into positive integers increases asymptotically
as \cite{as}

\begin{equation}
p(L) \sim \frac{1}{\sqrt{48}\,L} \exp \left\{ \pi \sqrt{\frac{2L}{3}} \right\},
\end{equation}

\noindent
the magnitude of the degeneracies becomes quite large, even for moderate
values of $L$. This set of states will be called the partition space of
$L$, and the problem of characterizing the yrast spectra thus amounts to
finding the lowest quantum states selected by  the contact interaction acting
within the partition space of $L$. \\

\noindent
{\bf Partition space}\\

\noindent
Since the construction of appropriate basis states in partitions space and their 
matrix elements play the crucial role in the subsequent discussion, it is
useful to begin with a brief discussion of some of the most elementary
features of this interesting space. The standard notation for a partition of
the integer $L$ is

\begin{equation}
\left( 1^{n_{1}} 2^{n_{2}}  \dots L^{n_{L}}\right),
\end{equation}

\noindent
where the $n_r$ are restricted by the condition

\begin{equation}
\sum^L_{r=1} r n_r = L.
\end{equation}

\noindent
The set of integers $\{ n_r \}$ satisfying (4) can thus be thought of as a set 
of quantum numbers characterizing a basis state in the partition space of $L$.

\noindent
In the present discussion we shall exploit two different ways of associating a 
basis state with a given partition $\{ n_r \}$.
The two realizations differ in their identification of the object that carries 
the $r$ units of angular momentum:\\

\noindent
{(\bf i)} \,\,$ r = m$, \,\,the address of a single particle state\\
\noindent
In this realization the $n_m$ count the number of different particles, each 
of which is carrying $m$ units of angular momentum.

\begin{equation}
\Psi_{\left\{n_{m}\right\}} = | (m = 1)^{n_{1}} (m = 2)^{n_{2}} \dots (m = L)^{n_{L}}\rangle.
\end{equation}

\noindent
{(\bf ii)} \,\, $r = \lambda$, \,\, the multipole order of a (collective) normal 
mode of excitation\\
\noindent
Assuming that the normal mode is excited by acting on the ground state with the 
collective operator $Q_{\lambda}$, we have

\begin{equation}
\Psi_{\left\{n_{\lambda}\right\}} \sim \Pi^L_{\lambda=1} 
\left(Q_{\lambda}\right)^{n_{\lambda}} \,\,| \,\, 0 \rangle .
\end{equation}

\indent
Although these two realizations each provide a complete set of states for 
describing partition space, these states can be very different when the
numbers $n_r$  of excitations is large; thus for a single mode we have the
overlap

\begin{eqnarray}
\langle (m = \lambda)^{n_{\lambda}} | 
\frac{1}{\sqrt{ n_{\lambda}!}} 
\left(Q_{\lambda}\right)^{n_{\lambda}} | 0 \rangle 
& = & \sqrt{\frac{N !}{\left(N-n_{\lambda}\right)!N^{n_{\lambda}}}}\\
\nonumber
& \sim & 
\exp{\left\{ - \frac{1}{2N }\left(n_{\lambda}\right)^2 \right\}},\\\nonumber 
\end{eqnarray}

\noindent
where we have employed the expression (8) for $Q_{\lambda}$. Thus for states involving 
condensates in which $n_{\lambda} \sim {\cal O}(L)$ where $L \gg \sqrt{N}$,
the two different states derived from the same partition become essentially
orthogonal.\\ \indent In thinking about the structure of the operators
$Q_{\lambda}$ it is fruitful to take inspiration from Feynman's arguments [6]
that identify the long wavelength, low energy excitations of superfluid helium
with modes that are excited by acting on the ground state with a sum of single
particle operators that impart the conserved quantum numbers (momentum in the
case of liquid helium, angular momentum in the present problem) to a single
particle of the condensate. Thus we shall assume

\begin{equation}
Q_{\lambda} = \frac{1}{\sqrt{2N \lambda !}} \sum^{N}_{p=1} z^{\lambda}_p ,
\end{equation}

\begin{equation}
z_p = x_p + iy_p ,
\end{equation}

\noindent
where the spatial coordinates are measured in units of the oscillator length. The operator 
(8) acting on the $L = 0$ many body ground state of the oscillator creates a
symmetrized state with angular momentum $L = \lambda$; the operator (8) raised
to the $n_{\lambda}$ power creates  the many body excitations that appear in
(6).\\ \indent The excitation energies of the normal modes (8) can be
evaluated by taking expectation values of the contact interaction in the
appropriate harmonic oscillator states

\begin{eqnarray}
\epsilon_{\lambda} & = &\langle 0 | Q^+_{\lambda} V Q_{\lambda} | 
0 \rangle - \langle 0 | V | 0 \rangle\\ \nonumber
& = & - v N \left(1 - \left(\frac{1}{2}\right)^{\lambda - 1}\right),
\end{eqnarray}

\noindent
where the first term in the parenthesis represents the loss of interaction 
energy due to the removal of a single particle from the condensate in the
ground state $(m = 0)$ of the oscillator, and the second term describes the
remaining interaction between the condensate particles and the excited single
particle moving in an orbit with angular momentum $\lambda$. The short range
of the interaction is responsible for the rapid decrease of the latter term as
the particle with increasing $\lambda$ moves in orbits with larger and larger
radius. It should be emphasized that the energies (10) will be somewhat
modified in a many mode excited state as a result of mode-mode interactions,
but these interactions contribute terms that are at most of order $v$ and
therefore small compared with the leading order term (10). Thus, at least for
$L \ll N$, we can consider the $Q_{\lambda}$ excitations as a gas of
non-interacting particles.\\

\noindent
{\bf Attractive interactions} $(v < 0)$\\

\indent
On the basis of the above, we are now in a position to  construct the yrast states 
and the low energy excitations as a function of the angular momentum $L$. The phase
diagram is somewhat simpler for attractive interaction than for repulsive and
so we begin with attraction. The mode energies (10) are seen to be positive
for all modes with the exception of $\lambda = 1$ which has $\epsilon _{1} =
0$; the vanishing of the mode energy for $\lambda = 1$ can be simply
understood from the fact that the operator $Q_1$ is proportional to the center
of mass coordinate and thus does not affect the interactions of the particles
which only depend on relative coordinates. Thus, in agreement with [3], we
find that  the yrast state is described by a wave function\\

\begin{equation}
\Psi_L \sim \left( Q_1\right)^L \Phi_0,
\end{equation}

\noindent
in which the excitation energy and angular momentum are entirely associated with the 
center of mass of the atom cloud moving around in the confining oscillator
potential while the relative coordinates and interactions remain exactly as in
the $L = 0$ state.\\ 
\indent The state (11) can be seen to involve two
distinct condensates: partly the usual Bose- Einstein condensate that puts all
the particles into the lowest state of the oscillator. This condensate is, in
the presence of a finite temperature, melted by excitations that move
particles to excited states in the oscillator and this results in a critical
transition at the Bose-Einstein condensation temperature [1]

\begin{equation}
T_{c} \sim N^{1/3} \hbar \omega_0.
\end{equation}

\noindent
In addition, the state (11) involves a condensate into a single mode ($\lambda = 1)$ 
out of the many degrees of freedom in partition space. The low energy degrees
of freedom of the partition space are described by excitations involving the
removal of particles from the $\lambda = 1$ mode and corresponding excitation
of the modes with $\lambda = $ 2, 3 4, $\dots L$. Such excitations imply a
melting of the partition space condensate and because of the (near) degeneracy
of all the $\lambda  \not= 1$ normal modes, the thermal average of their
contributions to the partition function can be be easily evaluated and yield a
critical temperature

\begin{equation}
T_{pc} = \frac{Nv}{lnL}
\end{equation}

\noindent
for the partition space phase transition. At this temperature
the specific heat has a singularity (decreasing by a factor of 2) and the
occupation of the $(\lambda = 1)$ mode decreases from the value $2L (1+
\sqrt{5})^{-1}$ to a value that is of order $\sqrt{L}$. This transition is, of
course, in a finite system and therefore the critical region is spread over a
finite interval in temperature. However this interval is

\begin{equation}
\delta T \sim \frac{Nv}{(lnL)^2}
\end{equation}

\noindent
and thus small compared to the critical temperature.\\

\noindent
{\bf Repulsive interactions} ($ v > 0$)\\

\noindent
The spectra resulting from repulsive interactions are obtained by inverting 
those discussed above and the structure of the yrast region involves an analysis 
of the configurations that would have been those of highest energy in the 
earlier discussion. It turns out that there are again two distinct condensates 
at each value of $L$ and two phase transitions as a function of temperature, and 
in addition there is a third type of phase structure giving rise to a quantal 
phase transition (i.e. a transition at $T = 0$) in which the nature of the 
condensate in partition space changes abruptly as a function of $L$. These 
different $T=0$ phases are distinguished by the different integer number of 
vortex lines that exist in the ground state. 

We can again, for repulsive interaction (and $L \ll N$), consider the spectra 
constructed from the collective excitations produced by the operators 
$Q_{\lambda}$, but now the expression (10) yields negative values. The fact that 
the excitations have a negative energy reflects the 
fact that for repulsive interactions the interaction energy in the $L = 0 $ state 
is a maximum and 
all the $Q_{\lambda} (\lambda \not= 1)$ excitations reduce the energy by putting 
nodes in the wave function for relative motion and letting the particles get a 
little further away from each other. The yrast states will therefore be obtained 
by a condensation into that mode that has the greatest energy gain per unit of 
angular momentum.  As can be seen from Table I, the quadrupole $(\lambda = 2)$ 
and octupole $(\lambda = 3)$ modes are optimal in this respect; thus, for $L
\ll N$,  the yrast state will be a condensate involving only these
two modes [8]. Excitations of the 
condensate are obtained by removing quanta from the condensate mode and placing 
them in other $Q_{\lambda}$ modes, always subject to the constraint (4). In the 
presence of a finite temperature such excitations deplete the condensate and 
lead to a phase  transition at a temperature of order (13).\\
\indent
As we go to higher angular momentum quite general arguments \cite{stringari} 
suggest that 
for repulsive interactions and $L \sim N$ the near yrast states will be 
dominated by a structure involving a unit vortex line
\begin{equation}
\Phi_1 = \left( \Pi^{N}_{p=1} z_p \right) \Phi_0.
\end{equation}

\noindent
In the state (15) all $N$ atoms have been collectively shifted outward
in the  radial direction of the ($x, y)$ plane effectively expanding the area,
reducing  the density, and thus reducing the energy associated with the short
range  repulsive interactions. 

\noindent
The state (15) represents a realization of the partition $(1^N)$ in the mode 
corresponding to (5) and as discussed in connection with equation (7)
represents a terribly complicated state if expressed in terms of $Q_{\lambda}$
 excitations.  Since there is no efficient way of combining a condensate in the 
$Q_2$ mode with a condensation in the  $m = 1$ single particle state (unit vortex),
the yrast state must undergo a violent rearrangement (many narrowly avoided crossings) 
at the critical angular momentum that separates these two regimes, ie a quantum phase
transition.

\indent 
In the region around $L=N$ the low energy degrees of freedom 
can be described by the $Q_{\lambda}$ operators acting on the vortex state,
but now the excitation energies $\epsilon_{\lambda}$  are different from those
give by (10) since the, $\lambda-excitation$ now represents a mode in which a
particle is removed from the condensate in $(m = 1)$ and excited to a state
with $m = \lambda + 1$. There is also an additional mode in these spectra that
results from the possibility of removing a particle from the $m=1$ condensate
and exciting it to the $m=0$ state. This $\lambda = -1$ mode can be excited by
the operator

\begin{equation}
Q_{-1} = \frac{1}{\sqrt{2N}} \sum^N_{p=1} \frac{1}{z_p}.
\end{equation}

\noindent
However the multiple excitations of this mode are not described by powers of the 
operator (16) since these powers contain terms that go outside the partition
space. Rather, the $n$th excitation of the $\lambda = - 1$ mode  is created by
the operator

\begin{equation}
Q^{(n)}_{-1} \sim \sum_{{p_{1}< p{_2} \dots <p{_n}}} \left(z_{{p}{_1}} z_{{p}{_2}} 
\cdots z_{{p}{_n}}  \right)^{-1}.
\end{equation}
The degree of freedom associated with the $\lambda = -1$ mode can be seen as a 
displacement of the vortex line perpendicular to the axis along which the 
angular momentum is aligned as described by the broken symmetry condensate wave 
function
\begin{equation}
\Psi \sim \left(\Pi^N_{p=1} \left( z_p - a \right)\right) \Phi_0. 
\end{equation}

\indent
The many interesting features of rotating Bose condensates that are briefly 
sketched in the present note, are the subject of a more systematic report that 
is under preparation in collaboration with Christopher Pethick. In this 
connection it should be mentioned that the rather special features of the 
confining potential assumed in the present note, do not appear to be crucially 
important; rather the qualitative features of the phase diagram with its 
multiplicity of phase transitions exhibit a considerable range of robustness  
under changes involving anharmonicity and anisotropy of the external potential.\\

\indent
I am indebted to my colleague Christopher Pethick in more ways than one: partly for 
his inspiring introduction to the Wonderful World of Bose Condensates in
external traps, partly for informing me of his results, obtained in
collaboration with L.P. Pitaevskii, on the structure of the ground state for
$L \not= 0$ and attractive interaction, but most of all for his posing the
seminal question, ``what is the structure of the yrast line for these
systems?". I have also received valuable help from A.D. Jackson in addressing
the combinatorial problems encountered in constructing quantum states in
partition space. I wish to thank Georgios Kavoulakis for pointing out an error
in the original manuscript for this article and for his collaboration in the
further development.
\\

\vspace{2cm}

{\small{
\begin{center}
Table 1\\
\vspace*{0.5cm}
\begin{tabular}{|c|r|r|}\hline
$\lambda $ & $ \epsilon_{\lambda}$  &$\epsilon_{\lambda}/\lambda$\\ \hline
1 & 0 & 0 \\
2 & - 1/2     & - 1/4\\
3 & - 3/4     & - 1/4\\
4 & - 7/8     & - 7/32 \\
5 & - 15/16   & - 3/16 \\\hline
\end{tabular}
\end{center}

\vspace*{0.5cm}
\begin{description}
\item[{\rm Table 1:}]
\vspace*{0.5cm}
Caption: \\
 Energies of phonon modes for repulsive interactions.\\
The energies (in units of $ N | v | $) are
obtained from (10), and apply to $Q_{\lambda}$
excitations in the yrast region for ${L\ll N}$.
\end{description}
}}


\begin{references}
\bibitem{} A detailed study, including the effect of mode-mode interactions
in the low angular momentum region has been carried out in collaboration with
G. Kavoulakis and C. Pethick. This work establishes the fact that in this
region the condensate involves only the quadrupole mode until a critical
angular momentum at which both octupole and quadrupole modes become involved.
\end{references}
\end{document}